\documentclass[prb,twocolumn,showpacs,preprintnumbers,amsmath,amssymb]{revtex4}

\usepackage{graphicx}
\usepackage{dcolumn}
\usepackage{bm}

\begin{document}

\title{Density Functional Study of Excess Fe in Fe$_{1+x}$Te: Magnetism
and Doping}

\author{Lijun Zhang}
\author{D.J. Singh}
\author{M.H. Du}

\affiliation{Materials Science and Technology Division,
Oak Ridge National Laboratory, Oak Ridge, Tennessee 37831-6114}

\date{\today}

\begin{abstract}
The electronic and magnetic properties of the excess Fe in
iron telluride Fe$_{(1+x)}$Te are investigated by density functional
calculations.
We find that the excess Fe occurs with valence near Fe$^{+}$, and
therefore provides electron doping with approximately one carrier per
Fe, and furthermore that the excess Fe is strongly magnetic. Thus it will
provide local moments that interact with the plane Fe magnetism, and
these are expected to persist in phases where the magnetism of the planes
is destroyed for example by pressure or doping.
These results are discussed in the context of superconductivity.
\end{abstract}


\maketitle

Recently, iron chalcogenides $\alpha$-FeSe and $\alpha$-FeTe,
another new family of Fe-based superconductors have been reported.
\cite{hsu,mizuguchi,yeh,kotegawa,liu,fang,li}
The superconducting transition temperature $T_{c}$
has increased from initial 8 K\cite{hsu} to 14\cite{fang} (15.2\cite{yeh})
K with appropriate Te substitution, and 27 K at high pressures (1.48 GPa).
\cite{mizuguchi}
While the presently known maximum
critical temperatures are lower
than in the Fe-As families,
\cite{kamihara,chen,zhian,rotter_Ba,wang_Li}
this binary system has drawn considerable attention due to the apparent
simplicity of the structure, the fact that it is As free, and the fact
that large crystals of Fe$_{1+x}$(Se,Te) can be grown enabling
detailed characterization by neutron and other measurements.
These compounds the $\alpha$-PbO structure, which consists of
a $c$-axis stack of FeTe sheets, with each sheet consisting a square planar
layer of Fe, tetrahedrally coordinated by Te, similar to the FeAs sheets of
LaFeAsO or LiFeAs. In fact from a structural point of view these
compounds are very similar to LiFeAs, with As replaced by a chalcogen
and the Li replaced by a site with a low partial filling of excess Fe.
According to literature, these compounds always form with excess Fe.
\cite{gronvold,Leciejewicz,chiba,finlayson,Fruchart,bao}

Electronic structure calculations for
the stoichiometric iron chalcogenides, Fe$X$,\cite{subedi}
show electronic structures and Fermi surface topologies
very similar to those of the other Fe-based superconductors.
\cite{singh-du,dong-j,yildirim,mazin,yin-prl,singh,ma-f}
There is a general proximity to magnetism, especially in FeTe, as well
as a substantially nested Fermi surface, which favors a spin density wave (SDW)
instability at the 2D ($\pi$,$\pi$) point.
While the mechanism for superconductivity in the Fe-based superconductors
is yet to be established, there is a strong association between the
occurrence of the SDW and superconductivity in the phase diagrams, with
superconductivity generally occurring when the SDW is destroyed either by
doping or by pressure.
The SDW is observed in most of the undoped Fe-As superconducting materials,
and is accompanied by a lattice distortion.
\cite{chen_sdw,delacruz,klauss,rotter_sdw,Ishikado}
For the chalcogenides,
a structural distortion with decreasing temperature was detected
in FeSe$_{(1-x)}$\cite{margadonna}
and superconductivity was found to be close to magnetic instability in
Fe(Se$_{(1-x)}$Te$_{x}$)$_{0.82}$ (the formula
does imply chalcogen vacancies but reflects excess Fe).
\cite{fang}
Furthermore, Fe$_{1+x}$Te is reported as magnetic, with properties depending
on stoichiometry in several older papers.
Bao {\em et al.}\cite{bao} based on neutron results
suggested a more complex incommensurate
antiferromagnetic order for the Fe(Se$_{(1-x)}$Te$_{x}$) system than in
the Fe-As based SDW phases.
On the theoretical side, magnetism driven by Se vacancies,\cite{lee}
non-collinear\cite{pulikkotil} and bi-collinear\cite{ma_bi}
antiferromagnetic state have been suggested.

Here we report supercell calculations investigating the role of the
excess Fe focusing on Fe$_{1+x}$Te.
We find that as might be expected, excess Fe donates charge
to the FeTe layers, acting as an electron dopant. Interestingly, it occurs
with a valence near Fe$^{+}$ with each Fe donating one carrier.
Furthermore, there is a very strong tendency towards moment formation
on the excess Fe. These moments will then interact with the magnetism
of the FeTe layers, perhaps complicating the magnetic order. They
would also be expected to persist into the regime where FeTe magnetism
is suppressed by doping or pressure, perhaps extending the range of
magnetic order in the phase diagram, and providing pair breaking in the
superconducting state.

\begin{figure}
\includegraphics[width=3.2in]{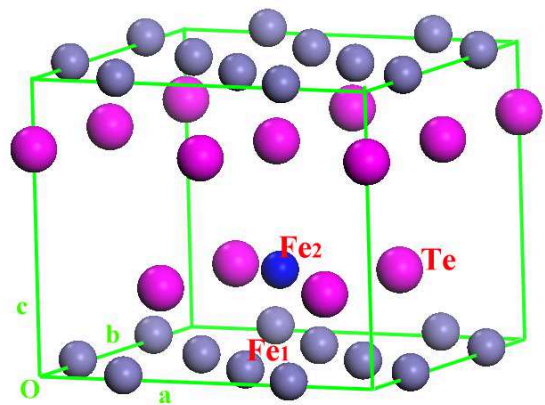}
\caption{(Color online)
Structure
(2x2 supercell of the tetragonal $\alpha$-FeTe with one excess Fe)
used to simulate Fe$_{1.125}$Te (composition Fe$_{9}$Te$_{8}$).
The iron in Fe-Te layers are denoted Fe$_{1}$ and excess iron as Fe$_{2}$.}
\label{str}
\end{figure}

The electronic structure and magnetism calculations were performed with the
projector augmented wave method\cite{kresse}
as implemented in VASP code.\cite{vasp1,vasp2}
The generalized gradient approximation\cite{pbe} was employed for the
exchange-correlation functional.
A kinetic energy cutoff of 268 eV and augmentation charge cutoff of 511 eV were
used to obtain converged energy (within 1 meV).
To simulate the partially occupied excess Fe,
we used a 2x2 supercell of $\alpha$-FeTe (two formula per cell)
with one Fe atom (labeled as Fe$_{2}$) placed at the 2$c$ (0.5,0,$z$) site,
as shown in Fig. \ref{str}.
This corresponds to a stoichiometry of Fe$_{1.125}$Te.
The experimental lattice parameters
$a$ = 3.8245 and $c$ = 6.2818 for Fe$_{1.125}$Te {\AA}\cite{Fruchart}
were used in our calculations.
An 8x8x10 grid was used for the $k$-point sampling of the Brillouin zone ,
and a denser 16x16x20 $k$-mesh was used for density of state (DOS) calculations.
The internal coordinates were relaxed to minimize the forces to below 0.01 eV/{\AA}. The calculated coordinate of Fe$_{2}$ is $z_{Fe}$ = 0.703. This is in reasonable agreement
with the experimental results in Ref. \onlinecite{Leciejewicz} ($z_{Fe}$ = 0.692) and Ref. \onlinecite{bao} ($z_{Fe}$ = 0.721), but significantly higher than that in
Ref. \onlinecite{Fruchart} ($z_{Fe}$ = 0.561).

\begin{figure}
\includegraphics[width=3.4in]{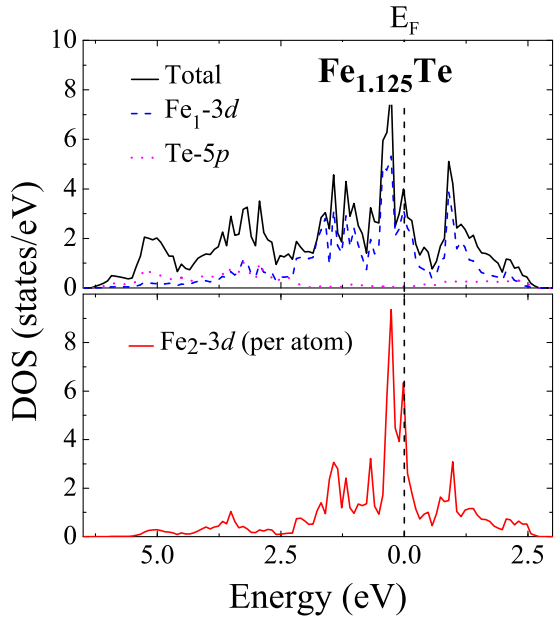}
\caption{(Color online)
Calculated electronic total and partial DOS for non-polarized Fe$_{1.125}$Te.}
\label{non}
\end{figure}

\begingroup
\squeezetable
\begin{table}
\caption{Calculated energy difference (in meV/Fe$_{1.125}$Te,
relative to the non-magnetic state for all of Fe)
between different types of magnetic arrangements
for Fe$_{1}$ in layers and excess Fe$_{2}$.}
\label{ene}
\begin{tabular}{l|cc}
\hline\hline
& Fe$_{2}$(non-magnetic) & Fe$_{2}$(magnetic)\\
\hline
Fe$_{1}$(non-magnetic) & 0 & -48.9\\
Fe$_{1}$(ferromagnetic) & -104.9\footnotemark[1] & -72.0\\
Fe$_{1}$(checkerboard antiferromagnetic) & -93.1 & -151.0\\
Fe$_{1}$(SDW antiferromagnetic) & -131.0 (-199.6\footnotemark[2]) & -186.7 (-256.6\footnotemark[2])\\
\hline\hline
\end{tabular}
\footnotetext[1]{Actually, this type of arrangement always converges to an ``antiferromagnetic'' order with Fe$_{1}$ and Fe$_{2}$ having opposite spin directions.}
\footnotetext[2]{The calculated value with relaxed structure when fully considering magnetic orders.}
\end{table}
\endgroup

\begin{figure}
\includegraphics[width=3.4in]{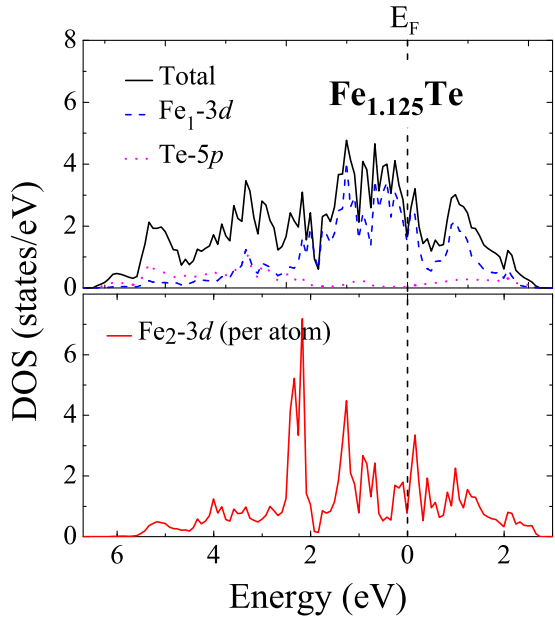}
\caption{(Color online)
Calculated electronic DOS for Fe$_{1.125}$Te with the moment formation on the
excess Fe$_{2}$ (non-magnetic order for Fe$_{1}$ layers).}
\label{part}
\end{figure}

We begin by showing that the excess Fe atom is strongly magnetic. Fig. \ref{non} shows the DOS for Fe$_{1.125}$Te obtained in a nonmagnetic calculation. The electronic states near the Fermi level ($E_{F}$) are mostly of 3$d$ character of the Fe$_{1}$ layers with small contribution from the excess Fe$_{2}$ atom. The result that the Fermi level lies exactly at a sharp peak of the Fe$_{2}$ 3$d$ DOS indicates the magnetic instability. The calculated Fe$_{2}$ partial DOS at $E_{F}$ is 6.2 states/eV/Fe (both spins). Within the Stoner theory the magnetism occurs when $N(E_{F})I$ $>$ 1, where $N(E_{F})$ is the DOS at the Fermi level per atom per spin and $I$ is Stoner parameter, typically in the range of 0.7 $\sim$ 0.9 eV for Fe. The large Fe$_{2}$ 3$d$ DOS at $E_{F}$ easily satisfies the Stoner criterion for the magnetic instability. Indeed, when considering the spin polarization for the Fe$_{2}$ atom (not Fe$_{1}$ layers), the total energy is reduced by 48.9 meV as shown in Table \ref{ene}. A pseudogap is opened with the Fermi level falling into its bottom, as shown in Fig. \ref{part}. The calculated Fe$_{2}$ DOS at the Fermi level is reduced to 0.8 states/eV/Fe. The calculated magnetic moment for Fe$_{2}$ is 2.5 $\mu_{B}$.

Based on
integration of the partial Fe$_{2}$ DOS up to the Fermi level and
normalization with the total Fe$_{2}$ DOS we find 4.7
electrons in the majority spin states and 2.2 electrons
in the minority spin states. Thus, the excess Fe occurs as Fe$^+$ and
each excess Fe atom has donates approximately
one electron to the Fe$_{1}$ layer. It may noted that Fe$^+$ is a somewhat
unusual valence state for stable Fe compounds. Here this state is stabilized
because of a balance between the two Fe sites. Specifically, in
stoichiometric FeTe, Fe is already di-valent, and so the more rapid
electron doping that would result if the excess Fe were di-valent would
lead to a more rapid conversion of the plane Fe towards Fe$^+$. This balance
between low valence states for Fe in the plane and excess positions may
be responsible for the fact that the structure does not form at
higher excess Fe concentrations.

For undoped FeTe, the Fermi level is located somewhat below the bottom of the pseudogap as shown in Ref. \onlinecite{subedi}. The presence of the excess Fe atoms in Fe$_{1.125}$Te moves the $E_{F}$ up, reducing the total DOS at $E_{F}$. However, despite the electron doping, the total DOS at $E_{F}$ remains relatively high (1.8 states/eV/Fe), which would still put the Fe$_{1.125}$Te close to magnetic instabilities. Our calculations show that stripe antiferromagnetic ordering (the SDW type) is most stable compared to the nonmagnetic, ferromagnetic, and the checkerboard antiferromagnetic phases (see Table \ref{ene}) assuming the fixed structure for the nonmagnetic phase in all these calculations. Relaxing the structure for the SDW antiferromagnetic state further lowers the total energy by nearly 70 meV.

The magnetic moment of the Fe$_{2}$ is calculated to be 2.4 $\mu_{B}$, much higher than that for the Fe$_{1}$ layers (1.6-1.8 uB). The excess Fe's strong magnetism is supported by the recent neutron scattering experiment.\cite{bao} The large local magnetic moment of the excess Fe is expected to persist even if the SDW antiferromagnetic ordering of the Fe layers is suppressed by the doping or pressure, thus causing pair breaking in the superconducting phase.

As we have found in our early work, FeTe has the strongest SDW state in the iron chalcogenide family.\cite{subedi} This is consistent with experimental
observations that
the SDW state is maintained up to
high excess Fe contents, which
as discussed correspond to high doping levels
(with $x$ up to 0.125 in the Fe$_{1+x}$Te systems).\cite{yeh,bao}
It should also be noted that the heavy doping corresponds to a large size
mismatch between the approximately cylindrical
electron and hole Fermi surfaces.
The driving force for an itinerant spin density wave is Fermi surface
nesting. As noted, \cite{subedi} for a large size mismatch the structure in
the susceptibility around (1/2,1/2) will develop a dip at the center, with
the maximum therefore moving off center. If the SDW stays stable the ordering
vector will then become incommensurate.
This apparently is the case
in Fe$_{1+x}$Te, and may explain the incommensurate SDW observed in
neutron scattering.
\cite{bao}
We note than an itinerant SDW can arise simply within this itinerant
framework, but would require a complex frustration within a local moment
picture.

Turning to the trends, as noted FeTe has a stronger tendency towards
magnetism than FeSe and the arsenides, but also still shows signatures of
spin fluctuations. \cite{subedi}
As such, within a scenario where superconductivity arises from pairing
due to spin fluctuations associated with the Fermi surface nesting, 
FeTe may have particularly high temperature superconductivity if the SDW
can be suppressed. However, the SDW persists up to high doping
levels. \cite{bao}
A particularly interesting experiment would be then to destroy the
SDW by pressure and search for superconductivity in the resulting
paramagnetic phase.
Furthermore, the fact that the excess Fe in this compound, and
presumably the Fe$_{1+x}$Se and Fe$_{1+x}$(Se,Te) superconductors,
has a local moment in proximity to the Fe layers offers an interesting
opportunity for experimental investigation of the interplay between
superconductivity and presumably pair breaking magnetic scattering
in the Fe superconductors.

In summary, we find that the excess Fe in Fe$_{1+x}$Te is strongly magnetic
and is also an electron donor, with
each excess Fe atom donating approximately one electron to the Fe layer.

\acknowledgements
We are grateful for helpful discussions with A. Subedi, I.I. Mazin,
D. Mandrus and B.C. Sales.
This work was supported by the Department of Energy,
Division of Materials Sciences and Engineering.

\bibliography{ExcessFe}
\end{document}